\documentstyle[12pt]{article}
\input psfig
\makeatletter
\def\section{\@startsection {section}{1}{\z@}{-3.5ex plus-1ex minus
    -.2ex}{2.3ex plus.2ex}{\reset@font\large\sc}}
\def\@maketitle{\newpage
 \null
 \vskip 2em
 \begin{center}%
  {\Large \@title \par}%
  \vskip 1.5em
   {\normalsize
   \lineskip .5em
   \begin{tabular}[t]{c}\@author
   \end{tabular}\par}%
  \vskip 1em
  {\normalsize \@date}%
 \end{center}%
 \par
 \vskip 1.5em}
\makeatother
\catcode`\@=11
\def\makepreprititle{\par
\begingroup
\def\thefootnote{\fnsymbol{footnote}}
\def\
@makefnmark{\hbox 
to 0pt{$^{\@thefnmark}$\hss}} 
\if@twocolumn 
\twocolumn[\@makepreprititle] 
\else \newpage
\global\@topnum\z@ 
\@makepreprititle \fi\thispagestyle{empty}\@thanks
\endgroup
\setcounter{footnote}{0}
\let\makepreprititle\relax
\let\@makepreprititle\relax
\gdef\@thanks{}\gdef\@author{}\gdef\@title{}
\gdef\@preprintnumber{}\gdef\@preprintdate{}\gdef\subtitle{}
\let\thanks\relax}
\def\preprintnumber#1{\gdef\@preprintnumber{#1}}
\def\preprintdate#1{\gdef\@preprintdate{#1}}
\def\subtitle#1{\gdef\@subtitle{#1}}
\def\@makepreprititle{\newpage
{\def\baselinestretch{1}
\begin{flushright} \@preprintnumber \par
\@preprintdate \end{flushright} } \par
\begin{center}
\vskip 1.5em
{\LARGE \@title \par} \vskip 2.5em 
{\large \lineskip .5em
\begin{tabular}[t]{c}\@author 
\end{tabular}\par}
\vskip 1em {\large \@date} \end{center}
\par
\vfil} 
\date{\sl Department of Physics, Tohoku University\\Sendai, 980 Japan}
\preprintdate{~}
\preprintnumber{~}
\subtitle{}
\def\endabstract{\if@twocolumn\else\endquotation\fi}

\catcode`\@=12

\def\lsim{\mathrel{\mathpalette\vereq<}}
\def\gsim{\mathrel{\mathpalette\vereq>}}
\catcode`\@=11
\def\vereq#1#2{\lower3pt\vbox{\baselineskip1.5pt \lineskip1.5pt
\ialign{$\m@th#1\hfill##\hfil$\crcr#2\crcr\sim\crcr}}}
\catcode`\@=12

\def\lsim{\mathrel{\mathpalette\vereq<}}
\def\gsim{\mathrel{\mathpalette\vereq>}}
\catcode`\@=11
\def\vereq#1#2{\lower3pt\vbox{\baselineskip1.5pt \lineskip1.5pt
\ialign{$\m@th#1\hfill##\hfil$\crcr#2\crcr\sim\crcr}}}
\catcode`\@=12

\begin{document}

\preprintdate{June 1994}
\preprintnumber{LBL-36175}
\title{Supersymmetry\thanks{This work was supported by the Director, Office of Energy 
Research, Office of High Energy and Nuclear Physics, Division of High
Energy Physics of the U.S. Department of Energy under Contract
DE-AC03-76SF00098.}\thanks{Invited talk presented at the 22nd INS
International Symposium on Physics with High Energy Colliders, Tokyo,
Japan, March 8--10, 1994, to appear in Proceedings of INS Symposium.
}}
\author{Hitoshi Murayama\thanks{On leave of absence from \sl Department
of Physics, Tohoku University, Sendai, 980 Japan}}
\date{\sl Theoretical Physics Group\\Lawrence Berkeley Laboratory\\
University of California\\Berkeley, CA 94720}
\makepreprititle

\begin{abstract}
I review phenomenologically interesting aspects of supersymmetry. 
First I point out
that the discovery of the positron can be regarded as
a historic analogue to the would-be
discovery of supersymmetry. Second I review the recent topics on the
unification of the 
gauge coupling constants, $m_b$--$m_\tau$ relation, proton decay, and
baryogenesis. I also briefly discuss the recent proposals to solve the
problem of flavor changing neutral currents. Finally I argue that the
measurements of supersymmetry parameters 
may probe the physics at the Planck scale.
\end{abstract}

\newpage
\renewcommand{\thepage}{\roman{page}}
\setcounter{page}{2}
\mbox{ }

\vskip 1in

\begin{center}
{\bf Disclaimer}
\end{center}

\vskip .2in

\begin{scriptsize}
\begin{quotation}
This document was prepared as an account of work sponsored by the United
States Government. While this document is believed to contain correct
 information, neither the United States Government nor any agency
thereof, nor The Regents of the University of California, nor any of their
employees, makes any warranty, express or implied, or assumes any legal
liability or responsibility for the accuracy, completeness, or usefulness
of any information, apparatus, product, or process disclosed, or represents
that its use would not infringe privately owned rights.  Reference herein
to any specific commercial products process, or service by its trade name,
trademark, manufacturer, or otherwise, does not necessarily constitute or
imply its endorsement, recommendation, or favoring by the United States
Government or any agency thereof, or The Regents of the University of
California.  The views and opinions of authors expressed herein do not
necessarily state or reflect those of the United States Government or any
agency thereof of The Regents of the University of California and shall
not be used for advertising or product endorsement purposes.
\end{quotation}
\end{scriptsize}

\vskip 2in

\begin{center}
\begin{small}
{\it Lawrence Berkeley Laboratory is an equal opportunity employer.}
\end{small}
\end{center}

\newpage
\renewcommand{\thepage}{\arabic{page}}
\setcounter{page}{1}

\section{Introduction}

Low-energy supersymmetry was introduced into particle physics in early
80's mainly because of a theoretical motivation, namely stabilizing
\begin{displaymath}
m_W^2 \ll M_P^2
\end{displaymath}
against radiative corrections \cite{naturalness}, where $M_P$ is the
Planck mass. This is still the main motivation for the theorists (at
least for me) to regard low-energy supersymmetry as a serious candidate
of physics beyond the standard model. If this idea is true, we expect
the discovery of supersymmetry in near-future collider experiments. This
would be really exciting and a historic event in particle physics.

Recently there has been a revival of interest in supersymmetry because LEP
measurements on $\sin^2 \theta_W$ \cite{LEP} have shown that the gauge
coupling constants extrapolated up to very high energies beautifully
meet at a single point using the supersymmetric particle content
\cite{meet}. More attention is now being paid to phenomenological
success of supersymmetry.

We should not believe in supersymmetry in a religious way, even after
the LEP measurement on $\sin^2 \theta_W$. Only experiments can decide
whether the world is supersymmetric. The aim of this talk is not to
convince the audience to {\it believe}\/ in supersymmetry. I am trying
to demonstrate how {\it interesting}\/ supersymmetry is, from the
theoretical, phenomenological, and cosmological aspects. In particular,
I wish to emphasize another virtue of supersymmetry in this talk,
namely, supersymmetry may offer us the unique possibility that the
measurement of supersymmetry breaking parameters at collider experiments
supplies us with a ``window to the Planck world'' \cite{ICEPP}.

\section{Motivation for supersymmetry}

It is often stated that the progress of elementary particle physics has
revealed an ``onion-like'' structure of microscopic substances. If
history repeats itself, then we may find another substructure of the
particles which appear to be elementary, especially in the Higgs sector
of the standard model. This understanding of the history leads to
speculations like preons or subquarks, or technicolor type scenarios.

However, I wish to remind the audience first that there is another
history which is often overlooked. A repetition of that history would
lead to a discovery of supersymmetry. It is the discovery of the
positron.

At the end of 19th century, there was a problem in electrodynamics that
the self-energy of the electron diverges. To see this, let us suppose
that the electron is a uniformly charged sphere with radius $r_e$. Then
a simple calculation shows that the electron has a self-energy due to
the Coulomb potential generated by itself,
\begin{equation}
E_{\it self} = \frac{3}{5} \frac{1}{4\pi \epsilon_0} \frac{e^2}{r_e}.
\end{equation}
The self-energy is linearly divergent in the point-like limit $r_e
\rightarrow 0$. This contribution can be depicted by a diagram where the
electron emits a photon and then re-absorbs it.

The observed mass of the electron is then a sum of the ``bare'' mass and
its self-energy, {\it i.e.},\/
\begin{equation}
m_e c^2 = (m_e)^0 c^2 + E_{\it self}.
\end{equation}
As we reduce the ``size'' of the electron, the smaller we should take
its ``bare'' mass $(m_e)^0$, maybe down to a negative value. It requires
increasing fine-tuning to reproduce the observed electron mass. Such a
theory cannot be true down to a small distance $r \lsim
\frac{1}{4\pi\epsilon_0} e^2/m_e c^2 \simeq 4$~fm \cite{LL}.\footnote{Let
me quote some sentences by Landau and Lifshitz \cite{LL}.

``Since the occurrence of the physically meaningless infinite self-energy
of the elementary particle is related to the fact that such a particle
must be considered as point-like, we can conclude that electrodynamics as
a logically closed physical theory presents internal contradictions when
we go to sufficiently small distances. We can pose the question as to
the order of magnitude of such distances. We can answer this question by
noting that for the electromagnetic self-energy of the electron we
should obtain a value of the order of the rest energy $mc^2$. If, on the
other hand, we consider an electron as possessing a certain radius
$R_0$, then its self-potential energy would be of order $e^2/R_0$. From
the requirement that these two quantities be of the same order, $e^2/R_0
\sim mc^2$, we find
\begin{displaymath}
R_0 \sim \frac{e^2}{m c^2}.
\end{displaymath}

``This dimension (the ``radius'' of the electron) determines the limit of
applicability of electrodynamics of the electron, and follows already
from its fundamental principles. We must, however, keep in mind that
actually the limits of applicability of the classical electrodynamics
which is presented here lie much higher, because of the occurrence of
quantum phenomena.''}
But of course we now
know the ``size'' of the electron is smaller than $10^{-3}$~fm!

The cure to this problem was supplied by the discovery of the positron.
The existence of the positron suggests that there is a fluctuation in
the vacuum where an electron positron pair is created and then
annihilated. Then another process is possible that an electron ``hits''
a positron created by the vacuum fluctuation and annihilates it, while
the other electron in the vacuum fluctuation remains and pretends it
were the original electron coming in. This gives us an intrinsic
uncertainty in the position of the electron of the order of the Compton
length $r_{\it Compton} \equiv \hbar/m_e c \simeq 400$~fm. Indeed, the
self-energy is cut-off at this scale due to a cancelation between two
processes (re-absorption and vacuum fluctuation) down to mild
logarithmic divergence \cite{Weisskopf},
\begin{equation}
E_{\it self} = \frac{3}{4\pi} \frac{1}{4\pi\epsilon_0} 
	\frac{e^2}{r_{\it Compton}} \ln \frac{m_e c r_e}{\hbar}
\end{equation}
where the ``size'' of the electron appears only in the log. Even for a
size equal to the Planck length, the self-energy is only about 10~\%
correction to the ``bare'' mass.

The cancelation of the linear divergence is a consequence of a new
symmetry, (softly-broken) {\it chiral symmetry}.\/\footnote{The chiral
symmetry is exact only in the limit where the electron is
massless. Though explicitly broken by non-zero electron mass, the chiral
symmetry prohibits the appearance of a huge self-energy, because the
breaking parameter $m_e$ is dimensionful and does not change the
short-distance behavior of the theory. Explicit breaking of a symmetry
which does not change the short-distance behavior of the theory is
called ``soft breaking.''} This symmetry transform electron to
positron.\footnote{More precisely, it transforms a positive energy
solution of the Dirac equation to a negative energy solution, and the
{\it absence} of an electron in the negative energy state corresponds to
a positron.} The discovery of a new symmetry lead to the cure of the
problem of the linear divergence in the electron self-energy.

The motivation for supersymmetry is very similar to the above
situation. If we consider the Higgs potential of the standard model and
calculate the self-energy of the Higgs field, it turns out to be
quadratically divergent. Therefore, the standard model with a naive
Higgs potential cannot be a true theory applicable to a length scale
much smaller than $10^{-17}$~cm. Though a fermion mass can be protected
by chiral symmetry as we have seen above, a scalar mass cannot be
protected by any symmetry of the scalar field alone. The way
supersymmetry cures this problem is as follows. First, chiral symmetry
protects fermion masses against quadratic and linear
divergences. Second, supersymmetry relates the scalar mass to the mass
of its superpartner, the fermion. The combination of these leads to only
a mild logarithmic divergence in the scalar mass. In diagrammatic
language, there is a cancelation between the diagrams of particles and
their superpartners.

To be more realistic, supersymmetry should be (softly) broken because we
have never observed a superparticle degenerate with the particles which
we already know. The effects of breaking can be characterized by the
supersymmetry breaking scale, $m_{SUSY}$. Then the self-energy
of the Higgs boson is roughly\footnote{This is in exact
analog with chiral symmetry. Chiral symmetry is softly broken by
the non-vanishing electron mass. The self-energy is proportional to the
explicit breaking.} 
\begin{equation}
\delta m_H^2 \sim \frac{\alpha}{4\pi} m_{SUSY}^2 \ln (m_H^2 r^2)
\end{equation}
where $r$ is the ``size'' of the Higgs boson. Since the Higgs mass
parameter is supposed to be around the $m_Z$ scale, we also
expect\footnote{Note that the suppression factor by coupling constants
is roughly compensated by a large log, if we take $r$ at the Planck
length.}
\begin{equation}
m_{SUSY} \sim m_Z.
\end{equation}

All these arguments are highly heuristic. For instance, another
possible cure for the problem of the quadratic divergence in the Higgs
mass is the replacement of the Higgs boson by a fermion pair bound
state. This is the idea of technicolor theories \cite{SF}. Though this
idea itself is very beautiful and intuitive, it is, unfortunately, hard
to implement fermion masses into this scenario; there is a competition
between realizing large enough top quark mass and small enough flavor
changing neutral current. There is the further logical possibility that
the ``bare'' mass of Higgs and the self-energy cancel to many digits to
give a Higgs mass at the 100~GeV scale. But I do not think we can rely
on such a theory as a framework to explain physics at its most
fundamental level. This is the point where individual taste comes into the
discussion. I do not think one can argue that supersymmetry is the only
possibility to solve this problem. Supersymmetry is one good candidate
which is known to be consistent with the present phenomenology. I will
concentrate on more phenomenological aspects of supersymmetry in the
next sections.

\section{Unification of gauge coupling constants}

\begin{figure}
\centerline{\psfig{file=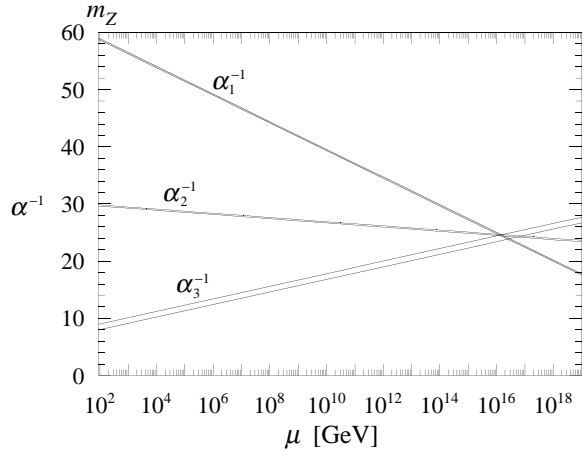,height=6cm}}
\caption[SU5]{The renormalization group evolution of the gauge coupling
constants with $m_{SUSY} = m_Z$.}
\label{SU5}
\end{figure}

Three gauge coupling constants are measured precisely at LEP/SLC
\cite{LEP,SLD}. If we extrapolate them to very high energies assuming
the particle content of minimal supersymmetric version of the standard
model, they beautifully meet at a scale $\simeq 2 \times 10^{16}$~GeV,
as everybody knows. This observation revived strong interest in
supersymmetric grand unified theories. The real excitement of this
observation is that the naive choice
\begin{displaymath}
	m_{SUSY} \sim m_Z
\end{displaymath}
is consistent with the unification of the gauge coupling constants (see
Fig.~\ref{SU5}).

After this simple but exciting observation, many people refined the
renormalization-group analysis to include threshold corrections both at
the GUT- and SUSY-scales \cite{EKN,RR,BH,HMY,HS,LP,HY,Yamada}. 

The questions asked in these works are the following.
\begin{enumerate}
\item How robust is the success of unification? Can it be destroyed by
threshold corrections?
\item Can we constrain $m_{SUSY}$ if we require unification?
\item Is the particle content of the MSSM unique to achieve the
unification? 
\item What can we learn about GUT-scale physics?
\end{enumerate}

Concerning the first question, it was found that the unification is in
general not destroyed even with threshold corrections. The threshold
corrections at the SUSY-scale are in general small. A simple
parametrization of $m_{SUSY}$ in terms of superparticle masses was
found, and in general $m_{SUSY}$ appearing in the renormalization group
analysis may be very different from the actual superparticle masses
\cite{LP}. Also, one has to take care of the difference of definitions
in $\overline{MS}$ and $\overline{DR}$ schemes where the latter
definition is ``more supersymmetric.'' Even the effects of the $SU(2)
\times U(1)$ breaking in superparticle masses have been discussed
\cite{Yamada}. In the end, SUSY-scale threshold corrections do not do
any harm to the unification of the gauge coupling constants.

Threshold corrections at the GUT-scale can in general be large. For
example, while the minimal $SU(5)$ model \cite{SU(5)} does not give us
large corrections, the missing partner model \cite{missing} which
includes large $SU(5)$ representations leads to a systematic difference
in the $\alpha_s$ prediction \cite{HY}. More complicated $SO(10)$ models
also may lead to big threshold corrections \cite{Sher}. These
corrections may somewhat weaken the beauty of unification, again they do
not destroy the unification.

The second question is a natural question especially in view of
near-future collider experiments. Unfortunately, the
renormalization-group analysis is only weakly sensitive to $m_{SUSY}$.
The dependence is only logarithmic with a small coefficient, and a naive
analysis shows that any $m_{SUSY}$ between $m_Z$ and 10~TeV equally well
leads to unification.  Furthermore, it was pointed out that the
GUT-scale threshold correction can completely destroy any more precise
prediction of $m_{SUSY}$ \cite{BH}. This is sad, but {\it c'est la
vie}.

The correct particle content is crucial in achieving unification of
gauge coupling constants. The MSSM particle content is usually assumed
in renormalization group analyses. It is interesting that the minimal
supersymmetric extension of the standard model leads to unification,
without the addition of arbitrary new fields into the model. On the
other hand, the particle content below the GUT-scale is now severely
constrained by the unification condition. For instance, the MSSM has two
Higgs doublets, while models with four doublets are in contradiction
with unification. In general, addition of gauge non-singlet fields
destroy unification.  There is still room for adding $SU(5)$ complete
multiplets to the MSSM particle content, since they change the slope of
the renormalization group running of the three gauge coupling constants
by the same amount. But again there is a severe constraint from the
requirement that the gauge coupling constants do not blow up below the
GUT-scale.  Introduction of one family and anti-family at the TeV scale
leads to the gauge coupling constants blowing up exactly at the
GUT-scale \cite{MMY}.\footnote{In this case, $\sin^2
\theta_W$ turns out to be at an ``infrared fixed point,'' so that its value
at $m_Z$ does not depend on the initial (large) values of gauge coupling
constants. Then one obtains the correct $\sin^2 \theta_W$ even without
grand unification.} Therefore, the particle content one can introduce at
the TeV scale is completely classified to be (i) ${\bf 5}^* + {\bf 10}$
(fourth generation), (ii) up to three pairs of ${\bf 5}+{\bf 5}^*$, and
(iii) ${\bf 10} + {\bf 10}^*$. Of course, the introduction of singlets
is completely harmless as far as the gauge coupling constants are
concerned.\footnote{It should be noted that the introduction of singlets
has a potential danger to destroy the hierarchy \cite{PS,KMY2}.}

An interesting possibility is that there is an enhanced gauge symmetry
below the GUT-scale, so that the contribution from new gauge multiplets
and new matter multiplets sum up to achieve the correct unification
\cite{DKR,KMY}. Though the unification is somewhat accidental in this
case, there are interesting aspects to these models. There is a right-handed
neutrino at an intermediate scale in accordance with the MSW solution
to the solar neutrino deficit as well as $\tau$-neutrino hot dark matter.

The last question is a very ambitious question, and one has to
completely specify one particular GUT model to answer it. The simplest
example is the minimal $SU(5)$ model \cite{SU(5)}. The new particle
content at the GUT-scale is an adjoint Higgs $\Sigma({\bf 24})$ and
Higgs quintets $H({\bf 5})$, $\bar{H}({\bf 5}^*)$, where the doublet
components of $H$, $\bar{H}$ are contained in the MSSM. Then the mass
spectrum at the GUT-scale can be parametrized by three quantities,
$M_V$, the mass of the heavy gauge fields corresponding to broken
generators of $SU(5)$, $M_\Sigma$, the mass of the adjoint Higgs, and
$M_{H_C}$, the mass of the color-triplet partner of the Higgs doublets
in $H$ and $\bar{H}$. What is interesting in this simple model is that
the current precision of LEP/SLC data can already constrain the mass
spectrum at the GUT-scale, on one combination $(M_V^2 M_\Sigma)^{1/3}$
and $M_{H_C}$ separately \cite{HMY}.

Using LEP data in 1992, one obtains \cite{HMY}
\begin{eqnarray}
&0.95 \times 10^{16}~\mbox{GeV} < (M_V^2 M_\Sigma)^{1/3} < 
	3.3\times 10^{16}~\mbox{GeV}&\\
&2.2\times 10^{13}~\mbox{GeV} < M_{H_C} < 2.3\times 10^{17}~\mbox{GeV}&
\end{eqnarray}
at a 90~\% confidence level. SLD data on $\sin^2 \theta_W$ prefers a lower
value of $M_{H_C}$ with central value down by a factor of 37, while new
$\alpha_s (m_Z) = 0.127 \pm 0.005$ from LEP prefers a larger value of
$M_{H_C}$. The constraint on $M_{H_C}$ is crucial to the
prediction of the proton decay rate, and it will be interesting to see what
values of $\alpha_s$ and $\sin^2 \theta_W$ the data converge to.

\section{$m_b$--$m_\tau$ relation}

The phenomenological success of SUSY-GUT is now not only the gauge
coupling constant unification, but also the $m_b$--$m_\tau$ relation.
Simple GUT models predict that $m_b$ and $m_\tau$ are the same at the
GUT-scale. The disparity between their observed masses is supposed to
arise from renormalization effects as we scale down from $M_{GUT}$ to
their mass shells \cite{Chanowitz}, in a similar manner to the gauge
coupling constants.

Recently there have been active discussions of the $m_b$--$m_\tau$
relation in the literature \cite{BBO}. The first observation is that the
successful unification of $m_b$ and $m_\tau$ at the GUT-scale requires
heavy top, or more precisely, large top quark Yukawa coupling to the
Higgs boson, namely $\simeq 1.1$ at the $m_Z$ scale. This does not
directly lead to a prediction of the top mass since there are two
Higgses in the MSSM, and we should introduce a parameter $\tan \beta$ to
relate the Yukawa coupling constant to the physical top mass. An
interesting point is that for any given $\tan \beta$ the top mass should
be nearly at its maximum possible value allowed by perturbation theory.
\begin{figure}
\centerline{\psfig{file=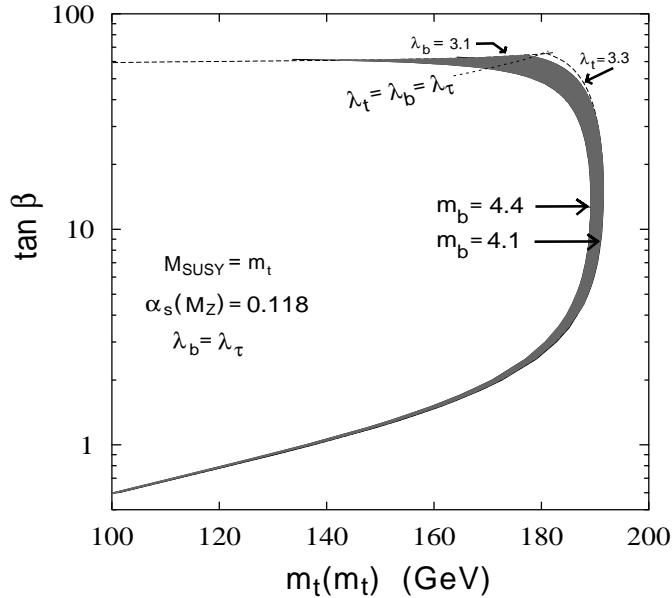,height=8cm}}
\caption[BBOO]{Constant $m_b$ contour in $(m_t, \tan\beta)$ plane
assuming $m_b=m_\tau$ at the GUT-scale \cite{BBOP}.}
\end{figure}

Further speculation is that even $m_t$ may be unified with $m_b$ and
$m_\tau$ at the GUT-scale. This is possible when $\tan \beta \sim 55$,
the smallness of $m_b/m_t$ being explained by the smallness of one Higgs
expectation value to the other. This led to a prediction of the top quark
mass since we know $\tan \beta$ in this assumption; typically $m_t
\simeq 180$~GeV \cite{HRS}. This looks consistent with the top mass
``evidence'' found by CDF \cite{CDF}! However, it was found later
that one needs to include a finite threshold correction for $m_b$ due to
superparticle loops \cite{HRS}, that weakened the predictive power.

To test whether $m_b$ and $m_\tau$ really unify at the GUT-scale, we
need to know $m_t$ more precisely, and also need to measure $\tan \beta$
by studying the Higgs bosons in the MSSM. Also in the large $\tan \beta$
region, we need to know (at least roughly) the superparticle spectrum.

\section{Proton Decay}

It is often stated that the non-SUSY $SU(5)$ GUT was excluded by the
proton decay experiments. It is true historically, but now it is an
empty statement because the gauge coupling constants do not meet in
non-SUSY $SU(5)$, so that one cannot predict where the GUT-scale
is. Therefore one first has to build models where the coupling constants
are unified to make predictions in non-SUSY GUT models
\cite{leptoquark}. I will not go into this direction in this talk.

There is another claim that SUSY GUT's do not have the same problem with
proton 
decay because the GUT-scale turns out to be much larger than that of
non-SUSY theories. This is true for proton decay induced by exchange of
heavy gauge bosons, leading to $p \rightarrow \pi^0 e^+$ as the
dominant mode. However, in most of the SUSY GUT models there is another
potentially more dangerous problem, namely proton decay via dimension-five
operators \cite{SKW}.\footnote{There is yet another problem on proton
decay by dimension-four operators, but they can be forbidden by a
discrete symmetry called $R$-parity.} Let us restrict ourselves to
the minimal SUSY $SU(5)$ model for a while.

Dimension-five operators are caused by the exchange of the color-triplet Higgs,
which is the $SU(5)$ partner of the doublet Higgs in the MSSM. Because of
supersymmetry, there is also a color-triplet Higgsino which is a fermion.
While exchange of heavy bosons induces operators suppressed by their mass
squared $\propto 1/M_{GUT}^2$, exchange of heavy fermions induces
operators suppressed only linearly in their masses, $\propto 1/M_{GUT}$.
Therefore, exchange of the color-triplet Higgsino is potentially a very
dangerous mechanism of rapid proton decay. 

Fortunately, dimension-five operators have very small Yukawa coupling
constants in front, since the color-triplet Higgsino couples to the
quark/lepton fields with the same strength as that of the doublet Higgs
in the MSSM because of $SU(5)$ symmetry. Furthermore, exchange of a
fermion cannot directly induce four-fermi operators which cause proton
decay. One needs to ``dress'' the dimension-five operators with a loop
of superparticles to obtain operators which are directly responsible for
proton decay. This gives us another small factor of $\alpha_2/4\pi$ or
so. In the end, the proton decay rate in the minimal SUSY $SU(5)$ model
turns out to be marginally allowed by the experiments, mainly Kamiokande
\cite{dimen5}. It is noteworthy that proton decay prefers a light
chargino (maybe observable at LEP200?) and heavy squarks (at the margin
of LHC reach).

\begin{figure}
\centerline{ \psfig{file=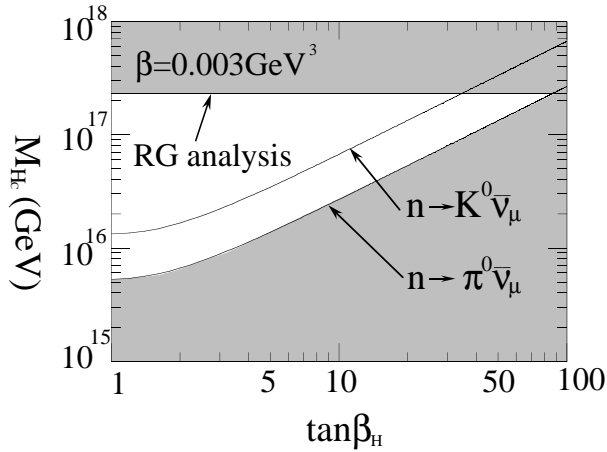,height=6cm} }
\caption[MHbound]{Constraints on $M_{H_C}$ from the renormalization
group analysis (from above) and the present nucleon decay experiments
(from below) in the minimal $SU(5)$ SUSY-GUT \cite{dimen5}. The shaded
regions are excluded at 90~\% C.L.  The horizontal axis is $\tan \beta$
of the MSSM Higgs sector.}
\label{MHbound}
\end{figure}

As I explained in section 3, LEP/SLC data are already sensitive to the
GUT-scale particle spectrum, and we have an {\it upper bound}\/ on the
mass of the color-triplet Higgs, $M_{H_C} \leq 2.3 \times
10^{17}$~GeV. On the other hand, proton decay (actually, neutron decay
can give better bounds in some cases) puts a {\it lower bound}\/ on
$M_{H_C}$, and it is interesting to see whether there is a remaining
region. The present lower bound is (see Fig.~\ref{MHbound})
\cite{dimen5}
\begin{equation}
M_{H_C} \geq 5.3 \times 10^{15}~\mbox{GeV},
\end{equation}
when taking the most conservative parameter set of hadron matrix element
{\it etc.}\/ 

As seen so far, proton decay in the minimal SUSY $SU(5)$ GUT is
predictive and can be probably tested by the superKamiokande
experiment. However, one may be suspicious of this model because it
gives us the wrong predictions $m_\mu = m_s$ and $m_e = m_d$ at the
GUT-scale. If we take the current quark masses determined by chiral
perturbation theory, more appropriate relations are $m_\mu = 3 m_s$,
$m_e = m_d/3$ \cite{GJ}. We have to modify the GUT-model to reproduce
the correct fermion mass spectrum. Then the prediction of proton decay
is also affected by the same modification, since the dimension-five
operators are induced by Yukawa interactions.  For example, the relation
$m_\mu = 3 m_s$ would increase the proton decay rate by a factor of four
compared to the minimal model. On the other hand, one can make a
complicated Higgs sector to kill dimension-five operators completely
\cite{BB}. Therefore, an improved lower bound on the proton lifetime
imposes stronger constraints on the SUSY-GUT models, but cannot exclude
them completely.

It is noteworthy that proton decay puts constraints even on the
physics at the Planck scale. If we allow arbitrary baryon-number
violating interactions at the Planck scale (as in some of the
superstring theories), they cause too-rapid proton decay. These
dangerous dimension-five operators can be forbidden using
``$B$-parity'' \cite{IR}, but this discrete symmetry is not consistent
with grand unification. One needs to resort on a flavor symmetry to
ensure the absence of dangerous operators. This puts strong constraint
on flavor physics model building \cite{KM}. In this way, an
improvement in proton decay experiments may lead to a deeper
understanding of flavor physics in the future.

\section{Neutralino Dark Matter}

One of the virtues of low-energy supersymmetry is that it leads to a
natural candidate for cold dark matter particle, the lightest
neutralino.

As explained in the previous section, one usually imposes a discrete
symmetry, $R$-parity, in the MSSM. This is a pure assumption, but seems
a plausible method to kill the dangerous dimension-four operators which
may cause too-rapid proton decay (with a life time of
$O(10^{-12})$~sec!).  $R$-parity is a very simple symmetry which assigns
even parity to all the fields of the standard model while odd parity to
their superpartners. An immediate consequence of this symmetry is that
the lightest superparticle (LSP) is stable. Then any LSP produced in the
early universe (at a temperature of $T \sim 10^2$~GeV) may possibly
remain until the present.

Actually, cold dark matter is needed anyway to explain structure
formation in the universe in a way consistent with the fluctuation in
the cosmic microwave background measured by COBE \cite{COBE}. For a
typical particle $\chi$ with a perturbative annihilation cross section, $\sigma
\simeq \pi \alpha^2/m_\chi^2$, its contribution to the present energy density
of the universe is $\Omega_\chi \sim 10^{-5} (m_\chi/\mbox{GeV})^2$. The
coincidence between the typical SUSY breaking scale $m_{SUSY} \sim m_Z$
and $\Omega_\chi \sim 1$ from this rough estimation is suggestive.

The precise calculation of the cosmic abundance depends on the details
of the superparticle mass spectrum. What is usually done is to assume the
``minimal supergravity'' boundary condition on the SUSY parameters at
the GUT-scale, which introduces five additional parameters to the
standard model. In this framework it turns out that the LSP is almost a
pure bino, the superpartner of the hypercharge gauge boson
\cite{LNY,DN}. A rough formula for the cosmic abundance is \cite{GKT}
\begin{equation}
\Omega_{\tilde{B}} \sim 
	\left( \frac{m_{\tilde{B}}}{60~\mbox{GeV}} \right)^{-2}
	\left( \frac{m_{\tilde{l}}}{140~\mbox{GeV}} \right)^4,
\end{equation}
and the requirement to avoid a charged dark matter $m_{\tilde{B}} <
m_{\tilde{l}}$ leads to an upper bound on superparticle masses,
$m_{\tilde{B}}, m_{\tilde{l}} \lsim 500$~GeV, or $m_{\tilde{g}} <
3$~TeV. This is an interesting constraint on the superparticle masses
without resorting on the naturalness arguments, though heavily relying
on the assumption of a bino-like LSP.

There are continuing active discussions/experiments relating to the
detection of the neutralinos in the halo in our galaxy
\cite{LBL}. Unfortunately, neutralinos have a very weak mainly
spin-dependent interaction with 
nuclei (weaker than the neutrino). We still have only a very weak limit on
neutralino dark matter from direct search experiments, while heavy
neutrino dark matter is almost excluded
\cite{Caldwell}. A recent topic is the discussion of constraints from
an indirect search at Kamiokande
\cite{kam}. Neutralinos in the galactic halo may be accumulated in the Sun or
Earth, and they may annihilate with each other to produce energetic
neutrinos. The neutrinos may then convert to energetic muons before
entering the neutrino \v{C}erenkov detectors to leave an ``upgoing
muon'' signature. Here the constraints are very sensitive to the SUSY
parameters, especially to the masses of the Higgs bosons.

\section{Baryogenesis}

Baryogenesis has been regarded as one of the main virtues of grand
unified theories \cite{Yoshimura}. There are many new particles at the
GUT-scale whose interaction violates baryon number. When they decay in the
very early universe, the decay may generate an asymmetry in
the baryon number by CP-violation. Unfortunately, the minimal $SU(5)$
SUSY-GUT does not share this virtue. The predicted baryon asymmetry
turns out to be too small because CP-violation in the decay of the
color-triplet Higgs appears only at two-loop order. One needs to extend
the Higgs sector of the model to begin with \cite{KTreview}.

Furthermore, there are general constraints on baryogenesis in the
SUSY-GUT models
which prefer new physics at an intermediate scale, rather than
generating the baryon asymmetry by a decay of a GUT-scale particle. The
first constraint is the monopole problem. Monopoles necessarily exist in
any grand unified theory based on a simple group \cite{Polyakov}. If
the universe starts at a very high temperature where the GUT-symmetry is
restored, monopoles are produced basically one for each horizon
\cite{Kibble}. Usually 
the annihilation of monopoles is negligible \cite{ZKP}. Then the energy
density of the universe is completely dominated by the monopoles, and
the universe is very short-lived; it turns back to a ``Big Collapse.''
One can avoid this problem if there is inflation \cite{inflation},
but the temperature after inflation (reheating temperature $T_{RH}$)
cannot be beyond the GUT phase transition temperature so as not to produce
monopoles again. Therefore it is not easy to produce GUT-scale particles
after inflation. The situation becomes even more severe in
supersymmetric theories. In supergravity, there exists a spin 3/2 partner
of the graviton, the gravitino, whose mass is supposed to be of the same
order as the other superparticles. The lifetime of the gravitino is very
long, $\sim 10$~minutes for $m_{3/2} \sim 1$~TeV. Then it decays {\it
after}\/ the nucleosynthesis producing many high-energy $\gamma$'s,
destroying the light elements. To avoid this, the number density of gravitinos
should be small, and this requires the reheating temperature $T_{RH}$ to be
smaller than $T_{RH} \lsim 10^{10}$~GeV \cite{gravitino}.  If we take
this constraint seriously, GUT-scale particles cannot play any role in
generating the baryon asymmetry.

On the other hand, it is now widely accepted that the standard model
itself violates baryon ($B$) and lepton ($L$) numbers at high
temperature due to the anomaly effect \cite{KRS}. Even if the decay of a
GUT-scale particle could generate a baryon asymmetry, it would be wiped
out except as far as there is a non-vanishing $B-L$ asymmetry. Therefore
it is necessary to generate a $B-L$ asymmetry at some stage in the early
universe.\footnote{There are two other possibilities discussed in the
literature. One is to generate baryon asymmetry at the electroweak phase
transition \cite{CKN}. The other is to employ some mechanisms to keep
baryon asymmetry even when $B-L=0$ \cite{preserve}.} However, simple
$SU(5)$ GUT models preserve $B-L$ symmetry and hence no asymmetry can be
generated. 

These problems can be cured just by adding right-handed neutrinos $N$ to
the MSSM particle content (MSSM+N) \cite{MSSM+N}. Then baryon asymmetry
is generated ``automatically'' irrespective of details in the
inflationary scenario.

We introduce the right-handed neutrino supermultiplets to the MSSM. The
superpotential of this model is
\begin{equation}
W_{\rm MSSM+N} = W_{\rm MSSM} +
	h_{ij} N_i L_j H_u + \frac{1}{2} M_{ij} N_i N_j,
\end{equation}
and the mass term breaks $L$ and $B-L$ invariance. I wish to remind the
audience that the existence of a right-handed neutrino is also preferred in
explaining the small neutrino mass in the MSW solution \cite{MSW} to the
solar neutrino problem via the seesaw mechanism \cite{seesaw}. The
neutrino mass is characterized by $\Delta m^2 \sim 10^{-5}$~eV$^2$
\cite{GALLEX}. If we assume that the MSW effect is due to the
$\nu_e$--$\nu_\mu$ oscillation and take an $SO(10)$-like ansatz for the
neutrino Yukawa matrix $h_{ij}$, one obtains $M\simeq
10^{10}$--$10^{13}$~GeV. This mass range has also a cosmological
interest since the $\tau$-neutrino mass turns out to be around
$m_{\nu_\tau} \sim 1$--$100$~eV, and can contribute to the hot dark
matter density of the present universe. Therefore, the MSSM+N
can naturally lead to a co-existence of the neutralino cold dark matter
and $\nu_\tau$ hot dark matter (mixed dark matter), which is favored now
by the observed spectrum of the density fluctuation from COBE to galaxy
clusters \cite{Primack}.

The important point is that the MSSM+N ``automatically'' generates a
lepton asymmetry if the mass $M$ of the right-handed neutrino is smaller
than the expansion rate during the inflation \cite{MSSM+N}. The scalar
component of the right-handed neutrino supermultiplet is driven to large
values at the end of inflation due to the quantum fluctuations in the
de~Sitter spacetime
\cite{quantum}. It oscillates after inflation, and decays. The decay
generates a lepton asymmetry via CP violation in the neutrino Yukawa
matrix. Finally electroweak sphaleron effects partially convert the
lepton asymmetry to a baryon asymmetry \cite{FY}. This scenario depends
on the assumption that there is an inflationary period, but does not
depend on the details of the inflationary models. Then everything occurs
without any further assumptions; hence an ``automatic'' scenario.

Actually one may be even more ambitious to expect that the scalar
component of the right-handed supermultiplet itself can drive a chaotic
inflation \cite{MSYY}. Then the simplest extension of the MSSM, namely
the MSSM+N can accommodate chaotic inflation, baryogenesis, mixed dark
matter, and solve monopole and gravitino problems. The only price one
has to pay is to assume that the quadratic potential $V = M^2
|\tilde{N}|^2$ persists beyond $M_P$ \cite{MSYY2}.

\section{FCNC Problem}

Though I have been presenting various phenomenological virtues of
supersymmetry in the previous sections, there is a big embarrassment due to
the existence of the superparticles below 1~TeV. The exchange of
squarks and gluino may lead to an unacceptably large flavor-changing neutral
current, especially in $K$--$\overline{K}^0$ oscillations. This
requires the squark masses to be highly degenerate at least for
$\tilde{d}$ and $\tilde{s}$ \cite{GM},
\begin{equation}
\frac{m_{\tilde{d}}^2 - m_{\tilde{s}}^2}{m_{\tilde{d}}^2}
	\lsim 6 \times 10^{-3} 
		\left( \frac{m_{\tilde{d}}}{\rm TeV} \right).
\end{equation}
We need an understanding of why squarks are so degenerate in mass.

The standard lore to explain the degeneracy is the following. Suppose
supersymmetry is broken in a ``hidden sector,'' which interacts with our
``observable sector'' of quarks and leptons only through gravitational
interactions. Let us denote the scale of supersymmetry breaking in the
hidden sector as $\Lambda_{SUSY}$.  The masses of squarks and sleptons
are generated by gravitional interactions with the hidden sector, at the
order of $\Lambda_{SUSY}^2/M_P$ or $\Lambda_{SUSY}^3/M_P^2$ depending on
the hidden sector models, and turn out to be the same for any scalar
particles because gravity is flavor-blind. Though this may sound
plausible, the supergravity Lagrangian does not have this feature in
general. No symmetry principle restricts the interactions between the
hidden and observable sectors to be flavor-blind.

Recently, there have appeared several interesting proposals to ensure the
absence of FCNC processes due to superparticle loops. I'll briefly
describe the basic ideas in the paragraphs below.

The first one is based on superstring theory (don't be afraid; I myself
am an amateur). In four-dimensional superstring models, there is a
so-called ``dilaton'' field which plays a unique role in superstring
theory. It has a completely flat potential to any finite order in
perturbation theory, but is supposed to have an expectation value due to
non-perturbative effects. This expectation value determines the gauge
coupling constants dynamically. The point is that the dilaton field has
a universal coupling to all fields. If the dilaton plays another role in
breaking supersymmetry, the squarks and sleptons acquire the same masses
at the Planck scale \cite{KL}. The assumption is that the dilaton has
two expectation values, one is the usual scalar expectation value which
determines the gauge coupling constants and the other is the so-called
$F$-component which breaks supersymmetry. Though this scenario is based
on relatively firm theoretical grounds, so far no concrete model of the
hidden sector is known which gives rise to an expectation value of the
$F$-component of the dilaton field.

Another possibility that has been pointed out is that the supersymmetry
breaking effects are fed to quarks and leptons by the gauge rather than
the ``gravitational'' interactions. Since the gauge interactions are
flavor-blind, the generated supersymmetry breaking terms would be also
flavor-blind. Though the idea sounds very simple, it actually requires a
drastic modification of supersymmetric models. First of all, the gauge
interactions are characterized by dimensionless coupling constants, and
the supersymmetry breaking masses of the squarks and sleptons are not
suppressed by $1/M_P$, rather $m_{SUSY} \sim \left(\alpha/4\pi\right)^n
\Lambda_{SUSY}$ with $n$ being model-dependent. Therefore this scenario
requires many new particles at multi-TeV energies. An explicit
realization was worked out recently \cite{Dine-Nelson}, which has a rather
complex structure with a symmetry group $SU(7)\times SU(2) \times
SU(3)_L \times SU(3)_R$ in addition to the standard model gauge
group. Attractive features are that there is no gravitino problem,
supersymmetry is broken dynamically, and the model can be embedded into
$SU(5)$ unification. The most unattractive feature is that the vacuum is
only a local minimum.

The two proposals which I have described above deal with the physics of
supersymmetry breaking to explain the smallness of the flavor changing
neutral currents. Other scenarios below take a different attitude with
emphasis on the flavor dynamics, the physics of the Yukawa coupling.

A simple solution to the FCNC problem is to assume a symmetry among
squarks of different generations to ensure the degeneracy of
squarks. This sounds a very natural idea, but it is in apparent
contradiction with the non-degeneracy of quarks. First of all, one needs
a non-abelian flavor symmetry so that (at least the first two
generations of) the quarks lie in an irreducible representation to
ensure $m_{\tilde{d}} = m_{\tilde{s}}$. Then it also restricts the form
of the Yukawa couplings. Therefore such a scenario should explain both
the observed structure of the Yukawa coupling matrix and the degeneracy
of squarks at the same time. One such example is based on a non-abelian
discrete group $\Delta (75)$ \cite{KS}. It is noteworthy that this
symmetry also prohibits dangerous dimension-five operators at the
$1/M_P$ level \cite{KM}. However one needs a relatively complicated
Higgs sector to break the flavor symmetry in the desired pattern. No
explicit Higgs potential has been presented so far.

A completely different direction is to give up the degeneracy of the
squarks, and try to explain the smallness of the flavor-changing neutral
current by yet another flavor symmetry. One of the reasons why an
exchange of the squarks can lead to large flavor-changing neutral
current processes is that the mass eigenstates of quarks and squarks
may be completely unrelated in the flavor space. One possible way to
suppress the flavor changing neutral currents is to align the mass
eigenstates of quarks and squarks \cite{NS}. This sounds miraculous, but
it was shown that a certain abelian horizontal symmetry can restrict the
form of both the Yukawa matrix and the squark mass matrix in such a way
that the diagonalization of both matrices can be done with almost the
same rotation. An unattractive point is that the assignment of
horizontal charges is rather ad hoc, fitted to explain the hierarchical
structure of the Yukawa matrix. Once one accepts it, however, the
alignment of the quark and squark bases is automatic.

Currently, there is no reason to choose one scenario over the others
except from the aesthetic point of view. We will, however, be able to
discriminate among them experimentally after we find the
superparticles. This will be the topic of the last part of my talk.

\section{Future Prospects}

First of all, supersymmetry predicts a relatively light Higgs boson. In
the MSSM, it has long been known that the lightest Higgs boson should be
lighter than the $Z^0$ \cite{Inoue}. Thus LEP-200 could have made a
definitive test of the MSSM.  Unfortunately, a large top quark Yukawa
coupling induces an important correction to this prediction at the
one-loop level. The upper bound is pushed up to around 130~GeV, well
beyond the reach of LEP-200 \cite{OYY}. We should recall that the
current LEP bound on the MSSM Higgs is not so tight: $m_h \gsim 45$~GeV
for standard $b\bar{b}$ mode and $\gsim 25$~GeV in the worst case when
it decays invisibly into a neutralino pair. There is still a wide
parameter space which LEP-200 will explore.

It is very nice that the proposed hadron facility, LHC, will cover
most of the parameter space of the MSSM Higgs sector.\footnote{It has
been known that there is a region in the parameter space of the MSSM
Higgs sector (``ozone hole'') which is very hard to be covered at LHC
\cite{KZ}. However, continuous efforts are being devoted to ensure the
covering of the whole MSSM parameter space at LHC \cite{Gunion-Higgs}.}
Recently there has been the interesting suggestion that the Tevatron
can search for a MSSM Higgs up to 120~GeV if a luminosity upgrade is
performed \cite{Mrenna}. 

What is more encouraging is that the upper bound on the lightest Higgs
mass does not change much even if we consider more complicated models,
with singlets, fourth generation, {\it etc.}\/ The mass of the Higgs
boson is proportional to (the square root of) the strength of the
self-interaction among the Higgs bosons, just like the mass of a fermion
is proportional to its Yukawa coupling. The self-coupling tends to
become stronger at higher energies due to the renormalization-group
running. If the self-coupling is too strong, it can happen that it
becomes infinite at some scale below the GUT-scale. Requiring that the
theory remains within the validity of the perturbation theory up to the
GUT-scale, we can put an upper bound on the strength of the
self-coupling, leading to an upper bound on the Higgs mass. At
tree-level, models with singlets have an upper bound of 150~GeV
\cite{Drees}. This bound is pushed further up by  one-loop effects,
but never beyond 180~GeV \cite{MO2}. The same is true for models with
heavy fermions beyond the top quark \cite{MO1}. An $e^+ e^-$ collider of
$\sqrt{s} = 300$~GeV can definitely exclude supersymmetric theories
based on the GUT idea \cite{Okada}.

Concerning the superparticles, the following characteristic of the mass
spectrum is important to future searches. Colored particles tend to be
heavy, while colorless particles are light. This is because the form of
the renormalization group equations implies that the gauge interactions
make the superparticle masses heavier. Colored particles have the
strongest gauge interactions and hence become the heaviest among the
superparticles. Typical mass ratio of the gluino to the lighter chargino
is about a factor of 4. Therefore, supersymmetry is an ideal target of
the high-energy experiments that $e^+ e^-$ and hadron colliders
literally play complementary roles. I refer further discussions to the
talk by Michael Peskin \cite{Peskin}.

Currently the most stringent bounds on supersymmetry come from LEP and
Tevatron experiments. LEP has excluded charginos below 45~GeV in a
model-independent way, while CDF excluded gluino masses below
100~GeV.\footnote{This bound depends weakly on an assumption of the MSSM
parameters.} These two bounds are comparable because of the
above-mentioned characteristics of the superparticle mass spectrum. A
Tevatron with Main Injector would extend the reach up to 300~GeV or so
with like-sign dilepton and $\not\!\!E_T$ when $m_{\tilde{g}} \simeq
m_{\tilde{q}}$ \cite{Baer-Tevatron}, while LEP 200 will find charginos
below 90~GeV. So far constraints from hadron and $e^+ e^-$ colliders
will improve more or less in parallel.

Meanwhile, HERA can search for the superparticles produced singly if
$R$-parity is broken. For instance, a squark below 270~GeV may be found
if it is produced as an $s$-channel resonance in $eq$ collision
\cite{Dreiner}.

To go beyond the reach of the present machines, we must wait until LHC
begins operation. Depending on parameters and analyses, LHC will extend
the reach of the gluino search up to 1.2--1.8~TeV
\cite{Baer-LHC,ATLAS}. This would basically cover the whole ``natural''
region of the superparticle mass spectrum.

In view of these numerous experimental programs in the near future, I
strongly expect the discovery of supersymmetry in a few to ten years. 

\section{Window to the Planck World}

Once supersymmetry is discovered, one may worry about losing one's
job. Actually it will be exactly the opposite. Since there are so many
superparticles which await detailed studies, it will be just the
beginning of a whole series of experiments. Indeed, measurement of
superparticle masses and mixings will give us information crucial to
distinguish between various GUT-models, scenarios of flavor physics and
supersymmetry breaking. In this sense, low-energy supersymmetry is a
messenger of the physics of the very high-energy scale to the scale
which is accessible by experiments.

First of all, I should emphasize that precision measurements of
supersymmetry parameters to several percent are possible at an
$e^+ e^-$ linear collider with a high beam polarization
\cite{Tsukamoto}. I refer to the talk by K.~Fujii \cite{Fujii}
concerning this point on which we worked together. We have demonstrated
that the measurements of the superparticle masses and cross sections at a
few percent level enable us to extract parameters in the supersymmetric
Lagrangian, for example, the mass parameters of $SU(2)_L$ and $U(1)_Y$
gauginos.\footnote{It was also pointed out that one can measure the
gluino mass at a $pp$ supercollider with a precision better than 10~\%
\cite{Gunion-gluino}. One has to assume the low-lying superparticle
spectrum to do the analysis. Presumably the combination of
data from $e^+ e^-$ on low-lying superparticle masses and from hadron
supercolliders on missing $\not\!\!E_T$ and multi-lepton signals would
allow us a measurement of the gluino mass at this precision.} 

\begin{figure}
\centerline{ \psfig{file=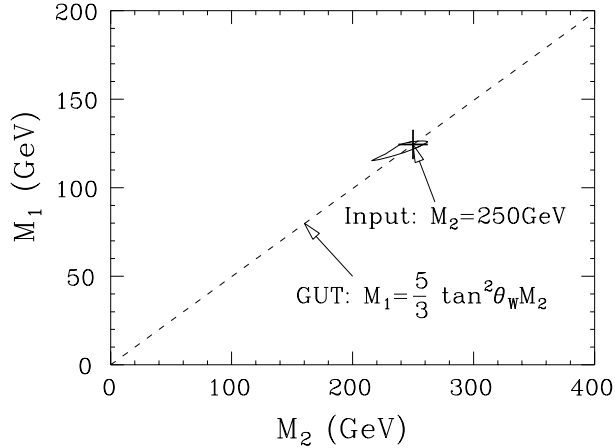,height=6cm,angle=90} }
\caption[jlcgut]{A $\chi^2 = 1$ contour in $(M_1, M_2)$ plane extracted
from chargino and selectron pair production processes at an $e^+ e^-$ 
linear collider after detector simulation \cite{Tsukamoto}. One can
test whether the gaugino masses unify at the GUT-scale with a 5~\%
precision.}
\label{jlcgut}
\end{figure}
There is a simple prediction in the SUSY-GUT models that the mass of the
gauginos should satisfy the relation
\begin{equation}
\frac{M_1}{\alpha_1} = \frac{M_2}{\alpha_2} = \frac{M_3}{\alpha_3}
\end{equation}
at any scale. We already know $\alpha_1$, $\alpha_2$ and $\alpha_3$
precisely thanks to LEP/SLC. They turned out to be consistent with
SUSY-GUT as we all know. The measurements of $M_1$ and $M_2$ would allow
us to make another test on SUSY-GUT models at the few percent level,
namely a test whether the gaugino masses unify at the same scale where the
gauge coupling constants unify (see Fig.~\ref{jlcgut}). It
would be even more exciting if the gluino mass measured at a hadron
collider also fits into the same relation. If the gaugino masses will be
consistent with the SUSY-GUT prediction, it would leave little doubt
about unification, at least of some kind.

The GUT-relation of the gaugino masses holds as far as the standard
model gauge group is embedded into a simple group, irrespective of the
symmetry breaking pattern \cite{KMY}. For instance, a model based on the
chain breaking (Pati-Salam)
\begin{equation}
SO(10) \rightarrow SU(4)_{PS} \times SU(2)_L \times SU(2)_R
	\rightarrow SU(3)_C \times SU(2)_L \times U(1)_Y
\end{equation}
predicts the same relation among the gaugino masses. Therefore, this
relation, if confirmed experimentally, would suggest a unification based
on a simple group,\footnote{It is noteworthy that the superstring with
dilaton $F$-term also leads to the same relation \cite{BIM}. This is
amusing because one needs rather big threshold corrections for the gauge
coupling constants to reconcile the difference between the apparent
GUT-scale and the string scale \cite{ILR}. Exactly the same correction
appears both for the gauge coupling constants and the gaugino masses to
give the same relation as in the (field theoretical) GUT models.} but
does not tell us about the existence of an intermediate scale or the
symmetry breaking pattern. On the other hand, models not based on a
simple group ({\it e.g.}\/, $SU(5) \times U(1)$ or superstring models
with moduli $F$-term) does not predict this relation.

Though all the GUT-models give the same relations for the gaugino
masses, the scalar mass spectrum distinguishes different models. Let me
discuss the ``vertical'' direction first, the spectrum within the same
generation. I will discuss the ``horizontal'' direction later, comparing
the masses of different generations.

\begin{figure}
\centerline{ \psfig{file=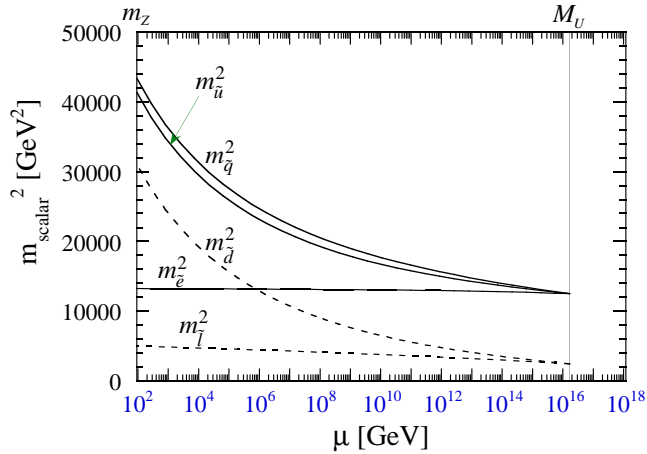,height=6cm} }
\caption[SU5scalar]{The renormalization group running of the scalar
masses in $SU(5)$ SUSY-GUT. They should unify at the same scale where
the gauge coupling constants unify \cite{ICEPP}.}
\label{SU5scalar}
\end{figure}

The difference of the masses for different quantum numbers reflects the
physics of the gauge symmetry. The first and the second generations have
tiny Yukawa interactions which can be neglected to a good
approximation. Then the splitting of the scalar masses within the same
generation arises only by their renormalization-group running due to the
gauge interactions. The pattern of the scalar mass spectrum will tell us
the pattern of the symmetry breaking. For instance, $SU(5)$ SUSY-GUT
predicts the scalar masses of $\tilde{e}_R$, $\tilde{u}_L$,
$\tilde{d}_L$ and $\tilde{u}_R$ unify at the GUT-scale in the same
manner as the gauge coupling constants unify (see Fig.~\ref{SU5scalar}).
Similarly for
$\tilde{e}_L$, $\tilde{\nu}_L$ and $\tilde{d}_R$. On the other hand, the
cases with an intermediate symmetry, such as Pati-Salam, predict certain
``sum rules'' of the scalar masses \cite{KMY},
\begin{eqnarray}
m_{\tilde{q}}^2 (M_{PS}) - m_{\tilde{l}}^2 (M_{PS})
        &=& m_{\tilde{e}}^2 (M_{PS}) - m_{\tilde{d}}^2 (M_{PS}),
                \label{PS1} \\
g_{2R}^2 (M_{PS}) (m_{\tilde{q}}^2 - m_{\tilde{l}}^2) (M_{PS})
        &=& g_4^2 (M_{PS}) (m_{\tilde{u}}^2 - m_{\tilde{d}}^2) (M_{PS}).
                \label{PS2}
\end{eqnarray}
at the scale $M_{PS}$ where Pati-Salam symmetry is broken. Though one of
the relations should be used to determine $M_{PS}$, the other relation
can be used to test the prediction. For models with lower symmetry,
predictivity becomes lower, so that they may not be able to be tested.
In any case, the scalar mass spectrum in the ``vertical'' direction
reflects the pattern of gauge symmetries at high energies. Table 1
summarizes the ``score sheet,'' of how different observables can
distinguish various models.
\begin{table}
\small
\centerline{
\begin{tabular}{lccc}
&$\alpha_i$ & $M_i/\alpha_i$ & $m^2_i$\\ \hline \\
$SU(5) \rightarrow G_{SM}$ & natural & common & testable\\
$SO(10) \rightarrow G_{SM}$ & natural & common & testable\\
$SO(10) \rightarrow G_{PS} \rightarrow
G_{SM}$ & adjustable & common & testable\\
$SO(10) \rightarrow G_{3221} \rightarrow G_{SM}$ & adjustable & common &
not testable\\ 
$SO(10) \rightarrow G_{3211} \rightarrow G_{SM}$ & adjustable & common &
not testable\\ 
$SU(5) \times U(1) \rightarrow G_{SM}$ & adjustable & common only for
$i=2,3$ & testable\\
superstring with dilaton $F$-term & adjustable & common & testable\\
superstring with moduli $F$-term & adjustable & not common & not testable
\end{tabular}
}
\caption[1]{The ``score sheet'' of how well we can distinguish between
various models \cite{KMY2}. The intermediate groups are defined as $G_{PS} =
SU(4)_{PS} \times SU(2)_L \times SU(2)_R$, $G_{3221} = SU(3)_C \times
SU(2)_L \times SU(2)_R \times U(1)_{B-L}$, $G_{3211} = SU(3)_C \times
SU(2)_L \times U(1)_Y \times U(1)_{B-L}$. The row $\alpha_i$ refers to
the unification of the gauge coupling constants, where ``natural'' means
that the unification is automatic, while ``adjustable'' employs either
particular particle content or threshold corrections to reproduce
observed gauge coupling constants. The row $M_i/\alpha_i$ refers to the
gaugino masses. The row $m^2_i$ states whether the model predicts a
definite pattern which is testable using the low-energy scalar mass
spectrum. }
\end{table}

The ``horizontal'' direction in the scalar mass spectrum carries
information on flavor physics at high scales. For instance, non-abelian
flavor symmetry gives us the degeneracy between the scalar masses of
different generations, for given gauge quantum numbers. Superstring
theories based on the dilaton $F$-term breaking also give the same
degeneracy, but with a further prediction of the ratio of the scalar
mass to the gaugino mass. On the other hand, a flavor group which leads
to an alignment of the quark and squark bases does not need a degeneracy
among the squarks, and we expect a baroque spectrum in the horizontal
direction.

In any case, the point is that we will be able to play the same game
with the superparticle masses as we play now with the gauge coupling
constants. We can make numerous checks whether experimentally
independent masses unify at higher energies. This is what I mean by 
``window to the Planck world'' \cite{ICEPP}. This is a truly
ambitious program, but would become possible if supersymmetry were true.

\section{Conclusion}

I reviewed interesting aspects of supersymmetry with emphasis on recent
topics. There are many more virtues and problems in supersymmetric models
which were not covered in this talk.\footnote{I refer to the ``Top-ten
list'' by Howard Haber \cite{Haber} for a more exhaustive list on both
virtues and problems of supersymmetry.} Though supersymmetry is
certainly a very interesting candidate of physics beyond the standard
model, we theorists do not know whether the nature is supersymmetric;
only experiments can decide it. And once supersymmetry is found, we will
gain many hints on the physics at ultra-high energy scales.

\section*{Acknowledgements}

I am grateful to all the organizers, especially to Prof. S.~Yamada, for
inviting me to give a talk in a well organized symposium. I also express
my sincere thanks to John March-Russell who gave me many useful comments
on the manuscript, and A.~Brignole who pointed out some mistakes.
Finally, I thank A.~Antaramian, A.~Brignole,
S.~Davidson, K.~Fujii, L.J.~Hall, J.~Hisano, D.~Kaplan, Y.~Kawamura,
M.~Luty, J.~March-Russell, T.~Moroi, Y.~Okada, K.~Olive, M.~Peskin,
A.~Rasin, R.~Rattazzi, H.~Suzuki, X.~Tata, T.~Tsukamoto, M.~Yamaguchi,
T.~Yanagida, and J.~Yokoyama for collaborations and fruitful
discussions.
This work was supported by the United States Department of Energy
under contract DE-AC03-76SF00098.


\begin{thebibliography}{99}
\bibitem{Mrenna}
 S.~Mrenna, talk presented at the preworkshop of SUSY 94, May 13,
 Michigan.
\end{thebibliography}
\end{document}